\documentclass[preprint,review,12pt]{article}

\usepackage{graphicx}
\usepackage{amssymb}
\usepackage{amsthm}
\usepackage{lineno}
\usepackage{amsmath}
\usepackage{algorithm}
\usepackage{algorithmic}
\usepackage{colortbl}
\usepackage{setspace}
\usepackage[normalem]{ulem}

\usepackage[top=3cm, bottom=2.0cm, left=2.5cm, right=2.5cm]{geometry}

\usepackage[numbers, comma, compress]{natbib}

\graphicspath{{Figures/}}

\newcommand{\vect}[1]{\boldsymbol{#1}}

\begin{document}

\begin{flushleft}

  \vspace{3cm}

  {\huge \textbf{Online Decorrelation of Humidity and Temperature in
    Chemical Sensors for Continuous Monitoring } }

  \vspace{1cm}
  Ramon Huerta$^{1}$,
  Thiago Mosqueiro$^{1}$,
  Jordi Fonollosa$^{2}$,
  Nikolai F Rulkov$^{1}$ and
  Irene Rodriguez-Lujan$^{3}$

  \vspace{0.3cm}
  {
  \small

  {\bf 1} BioCircuits Institute, University of California San Diego, La Jolla, USA

  {\bf 2} Institute for Bioengineering of Catalunya, Baldiri Reixac
  4-8, 08028 Barcelona, Spain

  {\bf 3} Dpto. de Ingenier\'ia Inform\'atica, Escuela Polit\'ecnica
  Superior, Universidad Aut\'onoma de Madrid, Calle Francisco Tom\'as
  y Valiente, 11, 28049 Madrid, Spain
  }

\end{flushleft}

\begin{abstract}
A method for online decorrelation of chemical sensor signals from the
effects of environmental humidity and temperature variations is
proposed. The goal is to improve the accuracy of electronic nose
measurements for continuous monitoring by processing data from
simultaneous readings of environmental humidity and temperature. The
electronic nose setup built for this study included eight metal-oxide
sensors, temperature and humidity sensors with a wireless
communication link to external computer. This wireless electronic nose
was used to monitor air for two years in the residence of one of the
authors and it collected data continuously during 537 days with a
sampling rate of 1 samples per second. To estimate the effects of
variations in air humidity and temperature on the chemical sensors
signals, we used a standard energy band model for an n-type
metal-oxide (MOX) gas sensor. The main assumption of the model is that
variations in sensor conductivity can be expressed as a nonlinear
function of changes in the semiconductor energy bands in the presence
of external humidity and temperature variations. Fitting this model to
the collected data, we confirmed that the most statistically
significant factors are humidity changes and correlated changes of
temperature and humidity. This simple model achieves excellent
accuracy with a coefficient of determination $R^2$ close to 1. To show
how the humidity-temperature correction model works for gas
discrimination, we constructed a model for online discrimination among
banana, wine and baseline response. This shows that pattern
recognition algorithms improve performance and reliability by
including the filtered signal of the chemical sensors.
\end{abstract}

\vspace{0.2cm} {\bf Keywords:} electronic nose, chemical sensors,
humidity, temperature, decorrelation, wireless e-nose, MOX sensors,
energy band model, home monitoring

\section{Introduction}

Conductometric chemical sensors are known to be very sensitive to
humidity levels in the environment
~\cite{barsan2001conduction,barsan2003understanding,Hubner2011347,morante2013chemical,buehler1997temperature,yamazoe2005toward,romain1997situ,hossein2010compensation,fine2010metal,oprea2009temperature,wang2010metal}.
This cross-sensitivity challenges the tasks of identification and
quantification of volatiles in uncontrolled scenarios. For example,
electronic noses can be used for human monitoring
purposes~\cite{romain2005monitoring,romain2008complementary,B307905H,ogawa2000monitoring,oyabu2001odor,oyabu2003proposition}.
In fact, they have been successfully used to quantify the number of
people working in a space-craft
simulator~\cite{fonollosa2014human}. In this case, it is likely that
the primary signal used by the algorithm to estimate the number of
people present at some given time is the humidity level in the
chamber. If we filter the sensor responses by the humidity and
temperature changes, a clearer chemical signature of the chamber can
be obtained, and this can facilitate more complex monitoring tasks
like identifying individuals \cite{rodriguez2013analysis}. A possible
solution to this sensitivity problem is the design of a special
sensing chamber that controls humidity and delivers the gas to the
sensors under predefined conditions
~\cite{chatonnet1999using,shevade2007development,ryan2004monitoring,fonollosa2014human,hossein2010compensation}.
Such preconditioning chambers are effective for signal improvement,
but their use increases the costs of electronic nose design for
applications in continuous monitoring of the environment
~\cite{B307905H}. A different approach is to build a model that
predicts the changes in the sensor conductance as a function of
humidity and temperature
variations~\cite{buehler1997temperature,hossein2010compensation,fort2011modeling,piedrahita2014next}.

The prevailing phenomenological model of sensor sensitivity is that
the ratio of the sensor resistance depends on a power law of the gas
concentration~\cite{windischmann1979model}. The model provides
accurate predictions when the gas is known and under controlled
conditions. However, it is rendered inaccurate with changes in the
environment. Correction methods based on artificial neural
networks~\cite{hossein2010compensation} using present and past values
of the input features are proven to be successful despite lacking an
explanation of the underlying processes. Fundamental models, on the
other hand, can capture the dynamical changes of resistance under
humidity variations accurately~\cite{fort2011modeling}. In these
models, the number of parameters is not large, but the model
parameters depend on the presented gas to the sensors.  Therefore, in
continuous monitoring systems, where there can be a complex mixture of
gases present in the air, it is indeed challenging to make proper
corrections on the sensor readings based on humidity and temperature
variations.

In this work, we propose an online methodology to subtract the changes
driven by humidity and temperature from the MOX sensor responses, and
demonstrate that this procedure enhances the performance of pattern
recognition algorithms in discriminating different chemical
signatures. We first develop a model based on the energy bands of
n-type semiconductors that is suitable for low-power micro-controllers
(Texas Instruments MSP430F247). We then make use of the predictions of
this model to subtract changes expected to be due to humidity and
temperature variation. Using a wireless electronic nose composed of 8
MOX sensors, we collected 537 days of data in the residence of one of
the authors and showed that our model is capable of predicting all MOX
sensors with a coefficient of determination $R^2$ larger than 0.9.
Because the electronic nose was subject to several unpredictable
conditions (house cleaning, wireless connectivity issues, etc), this
data set represents a wide variety of events present in home
monitoring scenarios. To evaluate the impact to online discrimination
of volatiles identities, we created a small data set consisting of
exposing the electronic nose to two distinct stimuli: wine and
banana. We show that the discrimination performance is significantly
enhanced using the decorrelated data combined with the raw time
series. This is a crucial task for any electronic nose system if one
wants to characterize or detect events based on their chemical
signatures in the presence of varying environmental conditions.

\begin{figure}[t]
 \centering
 \includegraphics[width=0.9\textwidth]{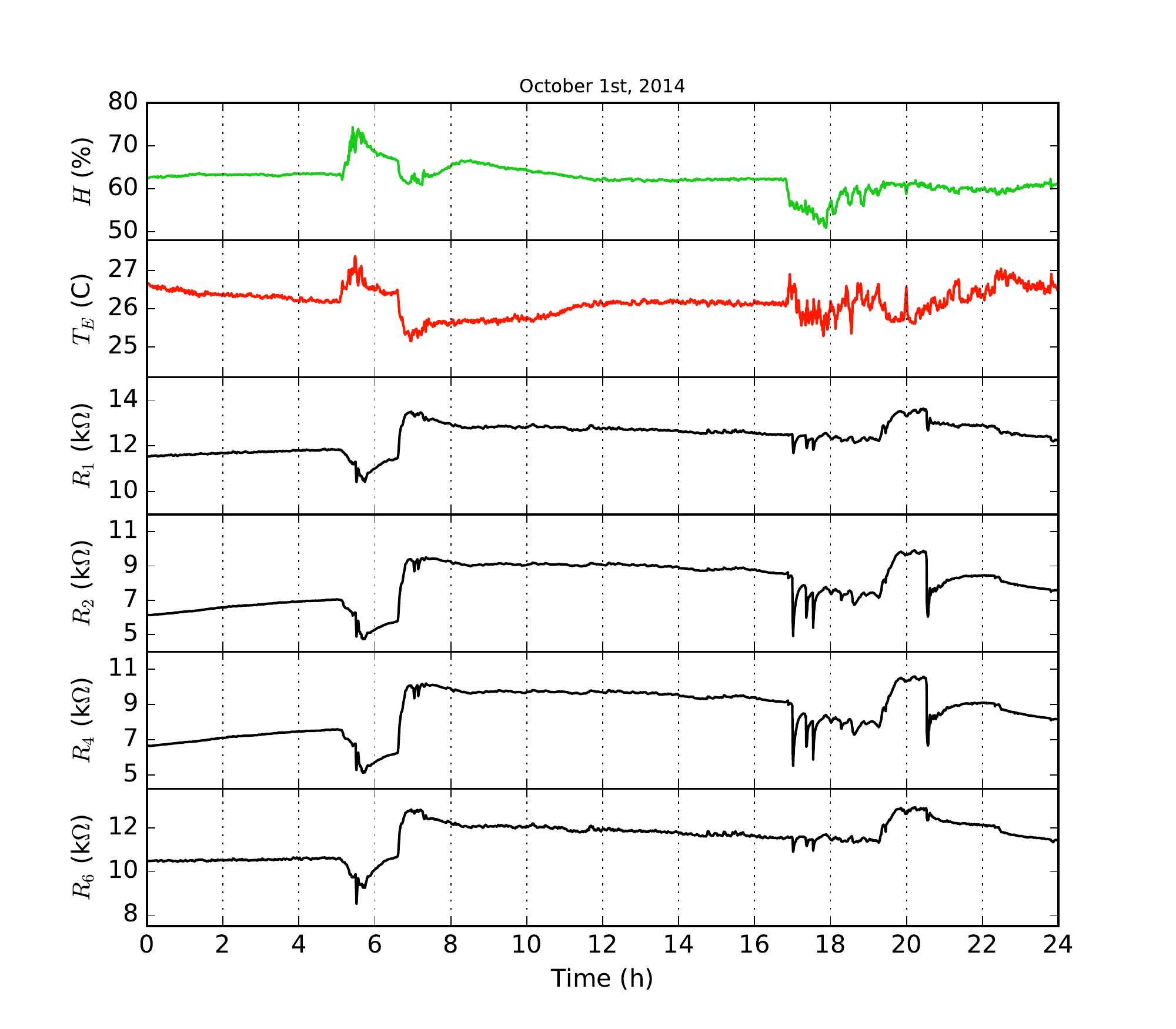}
 \caption{Illustrative example of recording during one day using the
   wireless electronic nose composed of 8 MOX sensors, including a
   humidity and temperature sensor. The first panel presents the
   humidity values, the second panel is the external temperature, and
   then resistance values for 4 different MOX sensors in the
   board. \label{fig:example} }
\end{figure}

\section{Example of sensors correlation with humidity and temperature}

In Fig.~\ref{fig:example}, we show a representative example of the humidity
problem using chemical sensors for continuous monitoring purposes. The
electronic nose in our setup is composed of 8 metal oxide (MOX) sensors, along
with temperature and humidity sensors. Such platform was previously used in our
wind tunnel studies to identify 10 gases at different
locations~\cite{vergara2013performance}. As a result of this previous
investigation, we know that we can discriminate between gases accurately, and
estimate gas concentrations in the ppm range ~\cite{fonollosa2014chemical}. The
time series shown in Fig. \ref{fig:example} were obtained in October 2014 in a
regular working day, in the residence of one of the authors.

The top panel shows the humidity levels throughout a complete day, where the
$x$-axis indicates the hour of the day. For example, the first rise in humidity
at about 5:30 AM corresponds to the morning shower. The sudden drop in humidity
at about 6:30 AM indicates opening the bathroom window, and the changes observed
at 5 PM are associated with the moment at which the family was returning home and the
door to the backyard was being opened. The second panel presents the temperature of
the electronic nose location that we denote by $T_E$ to differentiate it from
the temperature of the sensor heater, $T$. This residence did not have any air
conditioning system or heater operating during this period.

It is clear from this graph that the environmental changes in humidity
and temperature are often correlated. The measured resistance values
of the MOX sensors are presented in the four bottom panels. Although
the sensor board is made of 8 MOX sensors, here we present recordings
of only 4 of them because the remaining sensors are highly correlated
with those shown. Changes in the sensors resistance are strongly
affected by changes in humidity and temperature, as expected from the
extensive literature on the
topic~\cite{barsan2001conduction,barsan2003understanding,Hubner2011347,morante2013chemical,buehler1997temperature,yamazoe2005toward,romain1997situ,hossein2010compensation,fine2010metal,oprea2009temperature,wang2010metal}.
Nevertheless, the whole data set also includes examples where MOX
sensor changes cannot be explained only in terms of variations in
humidity and temperature as there also exist chemical variations in
the environment that have effects on sensors' responses.  As exposed
before, our goal is to find a way to decorrelate the MOX sensors from
humidity and temperature, and show that this improves pattern
recognition tasks such as discrimination of gas identity.

\section{Design of the wireless electronic nose}

\begin{table}[t]
\centering
 \begin{tabular}{@{}|l|c|l|}
\hline
Sensor type & Number of units &Target gases\\
 \hline
TGS2611&1&Methane\\
TGS2612&1&Methane, Propane, Butane\\
TGS2610&1&Propane\\
TGS2600&1&Hydrogen, Carbon Monoxide\\
TGS2602&2&Ammonia, H2S, Volatile Organic Compounds (VOC)\\
TGS2620&2&Carbon Monoxide, combustible gases, VOC\\
\hline \hline
\end{tabular}
 \caption{Sensor devices selected for the wireless electronic nose
   (provided by Figaro Inc.)}
\label{tablesensorts}
 \end{table}

In this section, we describe the electronic nose designed for home
monitoring purposes.  The sensor array is based on eight metal oxide
gas sensors provided by Figaro Inc. The sensors are based on six
different sensitive surfaces, which are selected to enhance the system
selectivity and sensitivity. Table 1 shows the selected sensing
elements along with the corresponding target compounds. In order to
control the variability between the sensing elements and increase the
flexibility of the sensing platform, the operating temperature of the
sensors can be adjusted by applying a voltage to the built-in,
independently reachable heating element available in each sensor. The
humidity and temperature sensors are integrated in the board using the
Sensirion SHT75. The device is very similar to the
M-Pod~\cite{piedrahita2014next}, except that ours is directly powered
by any electrical outlet to record continuously over long periods of
time.

\begin{figure}[t]
 \centering
 \includegraphics[width=0.6\textwidth]{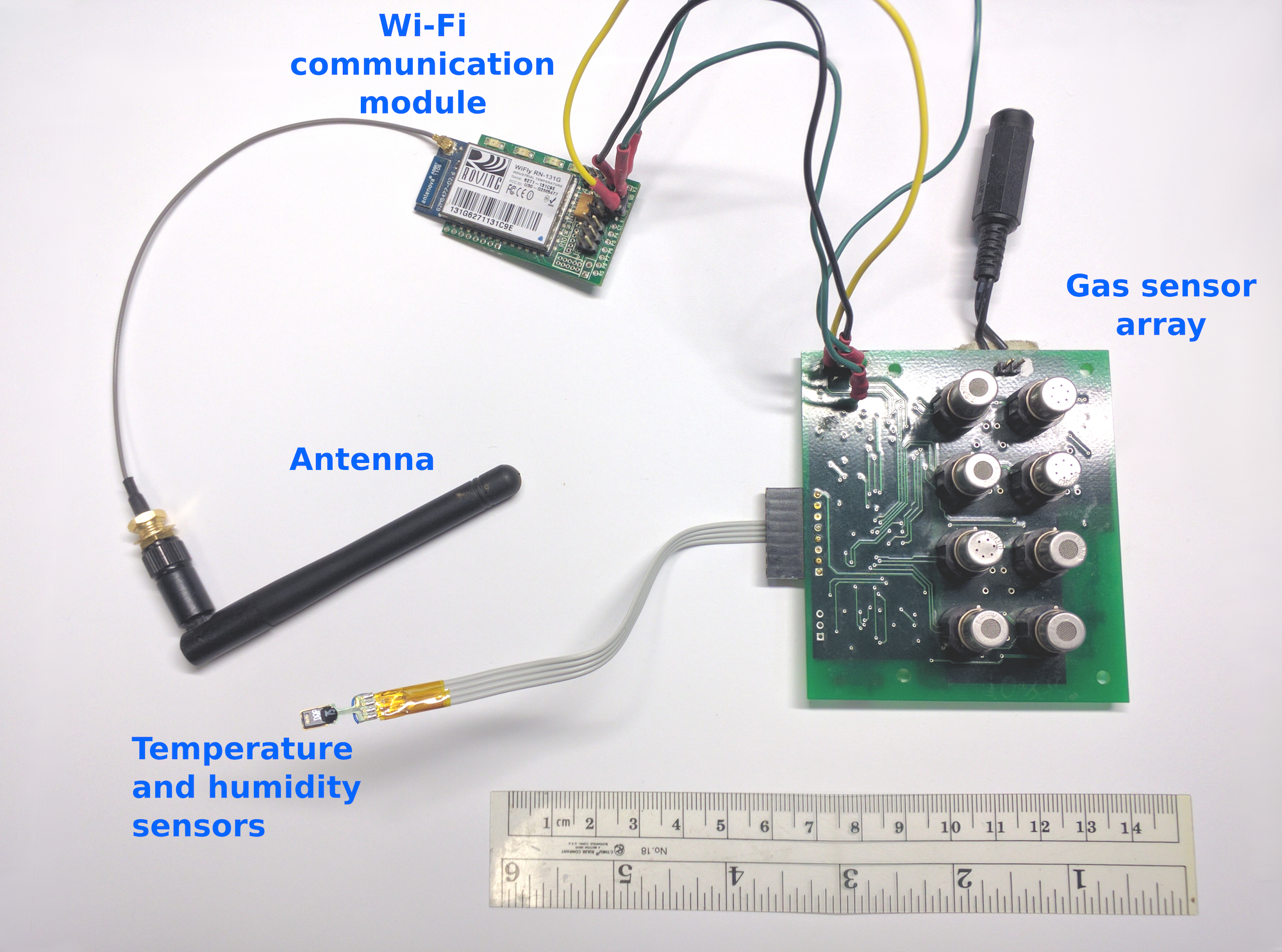}
 \caption{The electronic nose made of the sensor board (right) and a
   wireless communication board.\label{fig:enose}}
\end{figure}

The sensor array is integrated with a customized board that includes a
microprocessor MSP430F247 (Texas Instruments Inc.). In
Fig. \ref{fig:enose} we show the operating electronic nose. The
microcontroller was programmed to perform the following actions: i)
Continuous data collection from the eight chemical sensors through a
12-bit resolution analog-to-digital converter (ADC) device at a
sampling rate of 100 Hz; ii) Control of the sensor heater temperature
by means of 10 ms period and 6 V amplitude Pulse-Width-Modulated (PWM)
driving signals; iii) A two-way communication with another device to
transmit the acquired data from the sensors and control the voltage in
the sensors' heaters. The sensor board provides serial data
communication to another device via either a USB and/or a 4-pin
connector (Tx, Rx, Gnd, Vcc).

A wireless communication module acts as a bridge between the
MSP430F247 microcontroller and the network. The communication with the
MSP430F247 microcontroller is done via the UART port, whereas the
communication with the network is performed wirelessly. The board is
based on a WiFly RN-131G radio module included in a RN-134 SuRF board
(Roving Networks Inc). The WiFly module incorporates a 2.4GHz radio,
processor, full TCP/IP stack, real-time clock, FTP, DHCP, DNS, and web
server.

The module can be accessed via a RS-232 serial port (9600 default baud
rate) or a 802.11 wireless network so that its configuration can be
modified. The wireless communication module is configured such that it
accepts UDP and TCP connections, the baud rate of the microprocessor is
set to 115200 so that it can exchange data with the MSP430F247
microcontroller, and working with an external 4" reverse polarity
antenna to increase the power of the transmission.

\section{Online model for sensors response}
\label{sec:model}

An energy band model for n-type semiconductors describes the changes
in the resistance of the sensor before exposure, $R_I$, and after
exposure, $R_F$, as a nonlinear expression of the changes in the
semiconductor's energy
bands~\cite{barsan2001conduction,barsan2003understanding}. Energy
bands changes depend on variations in humidity and gas external
temperature, which modulates the overall transduction. If we denote by
$\Delta\Phi=\Phi_F-\Phi_I$ the work function change computed as the
difference between the work function after and before exposure, and we
express the electron affinity change as $\Delta \chi=\chi_F-\chi_I$,
the overall transduction can be expressed (following
~\cite{barsan2003understanding}) as:
\begin{linenomath}
  \begin{equation}
    \ln\left(\frac{R_F}{R_I}\right)=\frac{1}{k_B T}\left(\Delta\Phi-\Delta \chi\right),\label{eq1}
  \end{equation}
\end{linenomath}
where $k_B$ is the Boltzmann constant, and $T$ is the sensor operating
temperature controlled by the built-in sensor heater. The sensor
temperature is not constant because it is modulated by the external
temperature, $T_E$. To be able to build a basic model to be fitted to
the data, we make the following assumptions. We assume that
relative changes in the external humidity, $\Delta H= h$, and changes
in external temperature, $\Delta T_E=t$, are small enough. We
also assume that the chemical content
remains unchanged during the environmental changes. This assumption is important because it is known that humidity
changes induce nonlinear changes in the energy depending on the
chemical agent (see ~\cite{morante2013chemical}). Under these
assumptions, we can rewrite the transduction in equation~\ref{eq1} as
\begin{linenomath}
  \begin{equation}
    \ln\left(\frac{R_F}{R_I}\right)=\frac{1}{k_B (T+\mu
      t)}\left(\Delta\Phi(h)-\Delta \chi(h)\right),\label{eq2}
  \end{equation}
\end{linenomath}
where $\mu > 0$ is a dimensionless factor that reflects the impact of
the external temperature into the sensor.

Because the sensor board is based on a Texas Instruments MSP430F247
micro-controller, which can only perform simple mathematical operations, we now
consider in equation \ref{eq2} terms up to second order in $\Delta H$ and
$\Delta T$. This removes
most of the non-linearities from equation \ref{eq2}, but without oversimplifying
the model. We investigate the validity of this approximation in section
\ref{sec:performancemodel} in each of the sensors separately. Thus,
\begin{linenomath}
  \begin{multline}
    \ln\left(\frac{R_F}{R_I}\right) = \left(\frac{1}{k_B
      T}-\frac{\mu}{k_BT^2} t+\frac{\mu^2}{k_BT^3}
    t^2+O(t^3)\right)\times \\ \left(\Delta\Phi(0)-\Delta \chi(0)+\left[
      \left. \frac{\partial \Delta\Phi}{\partial h}\right|_{h=0}-
      \left. \frac{\partial \Delta\chi}{\partial h}\right|_{h=0}\right]
    h \quad + \right. \\ \left. \frac{1}{2}\left[\left. \frac{\partial^2 \Delta\Phi}{\partial
        h^2}\right|_{h=0}-\left. \frac{\partial^2 \Delta\chi}{\partial
        h^2}\right|_{h=0}\right] h^2 +O(h^3)\right) .
    \label{eq:app}
  \end{multline}
\end{linenomath}
Note that $\Delta\Phi(0)-\Delta \chi(0)=0$ because there are not
changes in humidity and temperature on our sampling time scale. The
simplified model is
\begin{linenomath}
  \begin{eqnarray}
    \nonumber \ln\left(\frac{R_F}{R_I}\right)&=&\frac{1}{k_B
      T}\left[\left. \frac{\partial \Delta\Phi}{\partial
        h}\right|_{h=0}-\left. \frac{\partial \Delta\chi}{\partial
        h}\right|_{h=0}\right] h + \frac{1}{2k_B
      T}\left[\left. \frac{\partial^2 \Delta\Phi}{\partial h^2}
      \right|_{h=0}- \left. \frac{\partial^2 \Delta\chi}{\partial
        h^2}\right|_{h=0} \right] h^2 \\ & &-\frac{\mu}{k_B
      T^2}\left[\left. \frac{\partial \Delta\Phi}{\partial
        h}\right|_{h=0}- \left. \frac{\partial \Delta\chi}{\partial
        h}\right|_{h=0}\right] h t \; .
    \label{prediction}
  \end{eqnarray}
\end{linenomath}
Therefore, we fit the following model to the data
\begin{linenomath}
  \begin{equation}
    \ln\left(\frac{R_F}{R_I}\right)=\beta_1 \Delta H+ \beta_2
    \left(\Delta H\right)^2+ \beta_3 \Delta H \Delta
    T_E,\label{linearmodel}
  \end{equation}
\end{linenomath}
where
\begin{linenomath}
  \begin{eqnarray*}
    \beta_1 &=& \frac{1}{k_B T}\left[\left. \frac{\partial \Delta\Phi}{\partial h}\right|_{h=0}-\left. \frac{\partial \Delta\chi}{\partial h}\right|_{h=0}\right] \\
    \beta_2 &=& \frac{1}{2k_B T}\left[\left. \frac{\partial^2 \Delta\Phi}{\partial h^2}\right|_{h=0}-\left. \frac{\partial^2 \Delta\chi}{\partial h^2}\right|_{h=0}\right] \\
    \beta_3 &=& -\frac{\mu}{k_B T^2}\left[\left. \frac{\partial \Delta\Phi}{\partial h}\right|_{h=0}-\left. \frac{\partial \Delta\chi}{\partial h}\right|_{h=0}\right] \; .
  \end{eqnarray*}
\end{linenomath}
Thus, our model has only three parameters to be fitted: $\beta_1$, $\beta_2$ and
$\beta_3$. In particular, $\beta_1$ and $\beta_3$ have opposite sign and they
are related by $\beta_3/\beta_1=-\mu / T$. This means that the ratio
$|\beta_3/\beta_1|$ becomes smaller with increasing sensor temperature.

\section{Results}
\label{sec:performancemodel}

\begin{table}[t]
  \centering
  \begin{tabular}{@{}|l|cc|cccc|}
    \hline
    Sensor & RMS & $R^2$  & $\beta_{1}\,(\beta_{1}/se(\beta_{1})$) & $\beta_{2}\,(\beta_{2}/se(\beta_{2})$) & $\beta_{3}\,(\beta_{3}/se(\beta_{3}))$ & $\beta_3/\beta_1$ \\
    \hline
    1 & 0.06 & 1.00 & -0.0044 (-128.14)$^*$& 0.00014 (38.40)$^*$& 0.0110 (58.41)$^*$ & -2.61 \\
    2 & 0.12 & 1.00 & -0.0110 (-186.04)$^*$& 0.00034 (54.11)$^*$& 0.0240 (71.75)$^*$ & -2.21 \\
    3 & 0.12 & 1.00 & -0.0110 (-187.12)$^*$& 0.00034 (53.57)$^*$& 0.0230 (69.60)$^*$ & -2.18 \\
    4 & 0.14 & 1.00 & -0.0110 (-190.95)$^*$& 0.00033 (55.31)$^*$& 0.0230 (73.06)$^*$ & -2.19 \\
    5 & 1.24 & 0.98 & -0.0056 ( -41.48)$^*$& 0.00018 (12.23)$^*$& 0.0086 (11.15)$^*$ & -1.54 \\
    6 & 0.48 & 0.99 & -0.0039 (-104.94)$^*$& 0.00012 (30.29)$^*$& 0.0071 (33.71)$^*$ & -1.84 \\
    7 & 2.06 & 0.90 & -0.0070 ( -99.24)$^*$& 0.00022 (28.94)$^*$& 0.0095 (23.57)$^*$ & -1.36 \\
    8 & 2.09 & 0.91 & -0.0057 ( -70.75)$^*$& 0.00020 (22.94)$^*$& 0.0029 ( 6.43)$^*$ & -0.52 \\
    \hline \hline
  \end{tabular}
  \caption{Results of fitting the model defined in equation (\ref{linearmodel}). The
  Root Mean Square (RMS) of the error in the predictions always remained below
  3.0, and the coefficient of determination $R^2$ was always above 0.9. We also
  show the coefficients $\beta_1$, $\beta_2$, and $\beta_3$ fitted for each
  sensor, along with their signal-to-noise ratio ($se(X)$ stands for standard
  error of $X$). All $\beta$ parameters are statistically significant
  (indicated with a *), with a $p$-value below $10^{-10}$.}
  \label{tableresults}
\end{table}

We fit the model defined in equation (\ref{linearmodel}) to data of 537 days
(from Feb 17, 2013 until June 5 2015) by down-sampling the time series to one
data point per minute and per sensor. Heaters for sensors 1-4 are always kept at
the same operating voltage, while sensors 5 to 8 are controlled under a protocol
that guarantees that the sensor responses always remain within a the same range
of values. Results summarized in Table~\ref{tableresults} prove the
effectiveness and statistical significance of the energy band model: the
accuracy rates achieved by the model, measured by the coefficient of
determination $R^2$, are above $90$\% for all sensors, and all the model
coefficients are statistically significant. Sensors with a fixed heater
temperature (i.e., sensors 1-4) outperformed sensors that operate with their
heater temperature actively changed (i.e., sensors 5-8). In the worst case
(sensor 8), the difference in $R^2$ is close to $10$\%. This probably suggests
that higher order terms become important in the approximation of equation
(\ref{eq:app}) when the heater temperature is actively changed. Moreover, as
predicted by equation (\ref{prediction}), the parameters $\beta_1$ and $\beta_3$
have opposite signs for all the sensors in the electronic nose. The ratios
$\beta_3/\beta_1$ estimated for the eight MOX sensors by our fitting (see
Table~\ref{tableresults}) are consistent with the voltage applied on the
sensors' heaters: obtained ratios for sensors 1-4 are similar as the sensors are
kept under the same heating conditions, and ratios for sensors 5-8 are lower as,
due to the active temperature control, they tend to be at higher temperature.

\begin{figure}
 \centering
 \includegraphics[width=1.0\textwidth]{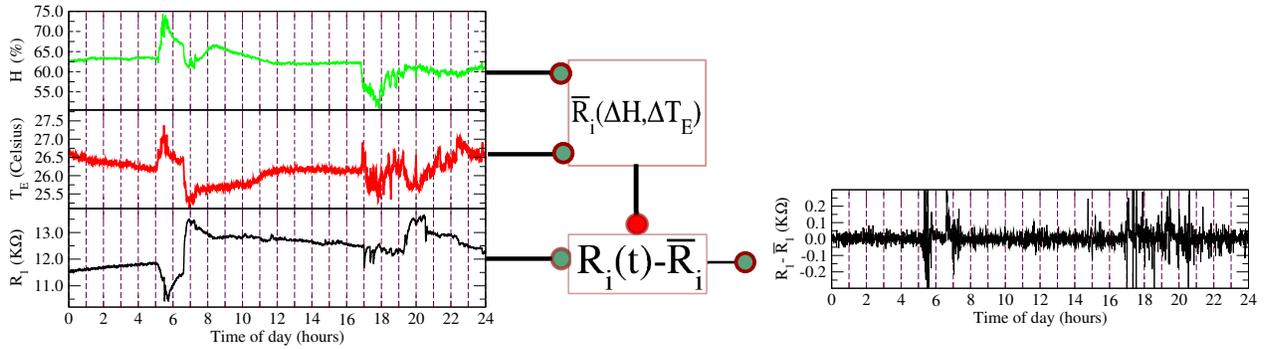}

 \caption{Result of applying the humidity and temperature filter provided by
 equation (\ref{linearmodel}) on sensor 1. First, the resistance is is predicted
 using the variation in humidity, and then this predicted resistance is subtracted
 from the original signal}
 \label{fig:filter}

\end{figure}

To filter the signal components due to changes in humidity and
temperature, we subtract the model prediction in equation~(\ref{linearmodel}) from the raw sensor
output. This operation is recognized as a method that searches signals
independent of environmental
conditions~\cite{cikeszczyk2013sensors}.
This is typically the case for continuous monitoring devices that are not
intended to measure the concentration of a particular gas. The resulting signal
is
\begin{linenomath}
  \begin{equation}
    R^*_i(t)=R_i(t)-\overline{R}_i(t)=R_i(t)-R_i(t-1)e^{\left(\beta_{1i}
      \Delta H+ \beta_{2i} \left(\Delta H\right)^2+ \beta_{3i} \Delta
      H \Delta T_E\right)},
    \label{eq:decorrmethod}
  \end{equation}
\end{linenomath}
where $R_i$ denotes the resistance values of the sensor $i$, and
$\beta_{1i}$, $\beta_{2i}$, and $\beta_{3i}$ are the adjusted values
for $\beta_1$, $\beta_2$, and $\beta_3$ for the i-th sensor. In
Fig. \ref{fig:filter}, we show the result of applying this
transformation on sensor 1. On the left panel, we present the
humidity, temperature, and sensor output. After applying the
transformation, the decorrelated output of the sensor is shown on the
right panel. The sensor drift due to the temperature and humidity
changes is filtered out. However, because we are subtracting from the
sensors signal ${R}_i(t)$ their predicted value $\overline{R}_i(t)$
according to our model, the resulting filtered signal $R^*_i(t)$ often
has zero mean and the relationship among the sensors is partially
lost. This is important for gas discrimination
\cite{vergara2013performance}, and we deal with this issue in section
\ref{sec:performance}.

\subsection{Parameter Stability}

\begin{figure}[t]
 \centering
 \includegraphics[width=0.5\textwidth]{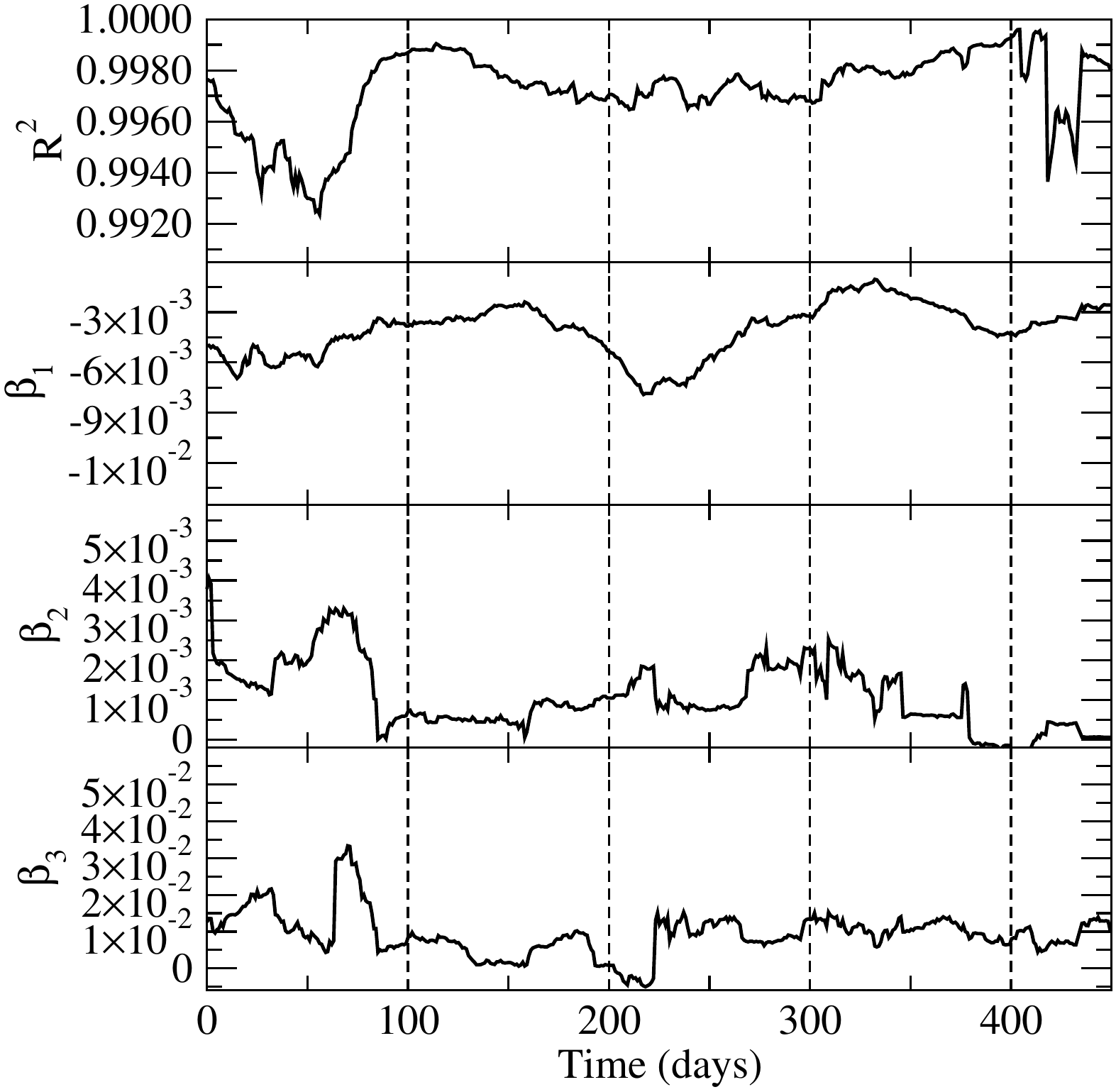}
 \caption{Time evolution of the out-of-sample performance measured by
   evaluating $R^2$ on the first sensor of the electronic nose. The
   three bottom panels represent the evolution of the parameters,
   $\beta_1,\beta_2$ and $\beta_3$ of the model over time.}
 \label{fig:stability}
\end{figure}

\begin{figure}[t]
 \centering
 \includegraphics[width=0.7\textwidth]{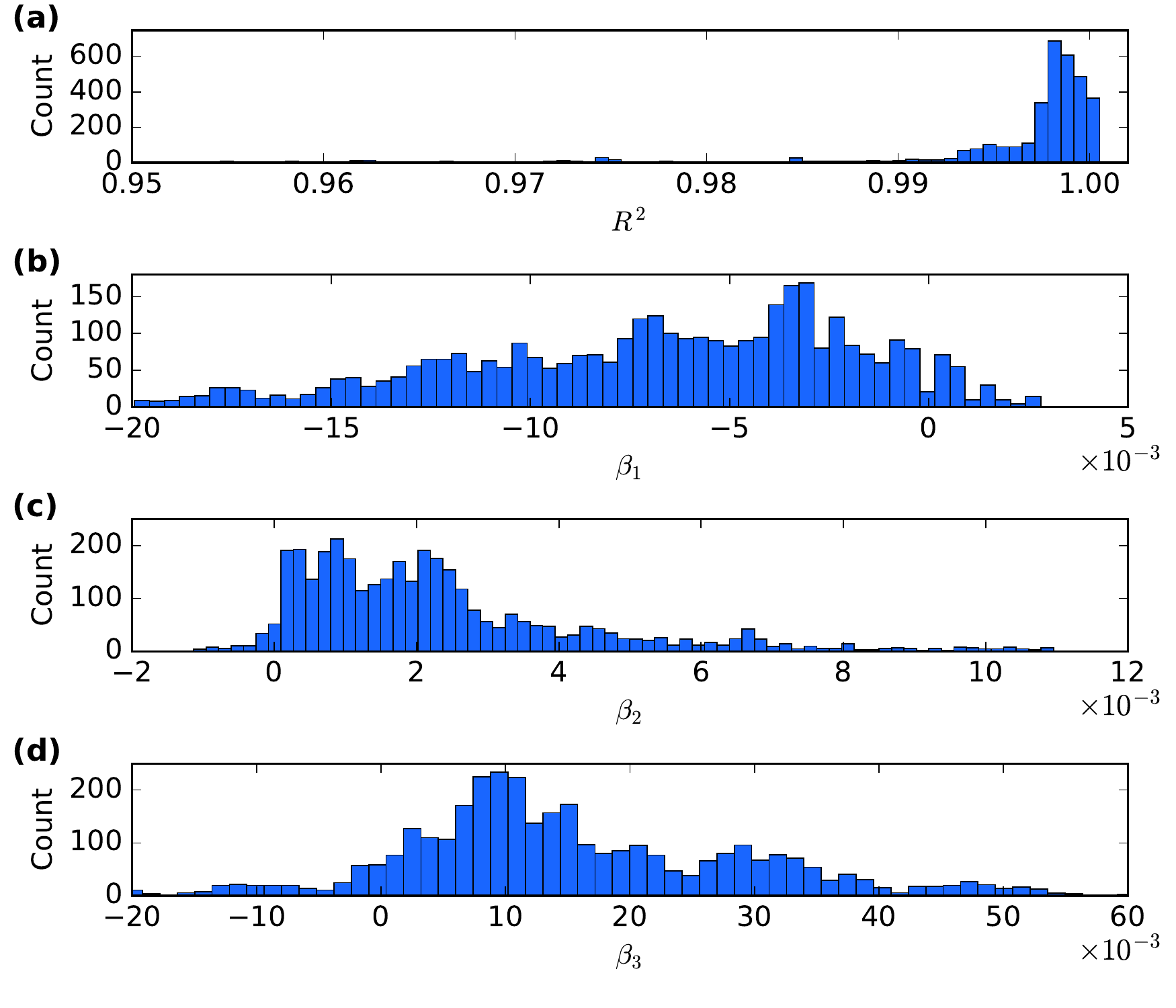}

 \caption{Histograms of performance $R^2$ (a) and values of $\beta$ parameters
 (b-d) for all the sensors using 3 months of training and testing in the
 following month.}

 \label{fig:stability2}
\end{figure}

To test the stability of the parameters over time, we trained the
model over a short period of time of 3 months of data and tested its
performance in the following month (i.e., forward testing
methodology). In Fig. \ref{fig:stability}, we show the time evolution
of the model performance and parameters $\beta$ of sensor 1 based on
humidity and temperature changes.  The window of 3 months was chosen
in order to guarantee $R^2 > 0.9$ for all sensors throughout the year
(Fig. \ref{fig:stability2}a) and to avoid longer time scales, where
sensor drifting and seasonal changes in the environment may influence
sensors response. We also show the histogram of all values assumed by
$\beta$ parameters throughout this period
(Fig. \ref{fig:stability2}b--d).

Finally, the model is robust to failures in the sensors due to number
of reasons. For instance, in some instances the electronic nose
stopped transmitting due issues in the wireless connectivity; in other
events, sensors were displaced from their location during house
cleaning, and stopped working. Because algorithms need to be as robust
as possible given the uncontrolled conditions under which they
operate, our $R^2$ already takes it into account. In summary, there
are many possible reasons in daily operations that hinder the
operation of the electronic nose, and they reproduce uncontrolled
conditions that such sensors face.

\subsection{Sampling rate}

\begin{figure}
 \centering
 \includegraphics[width=0.7\textwidth]{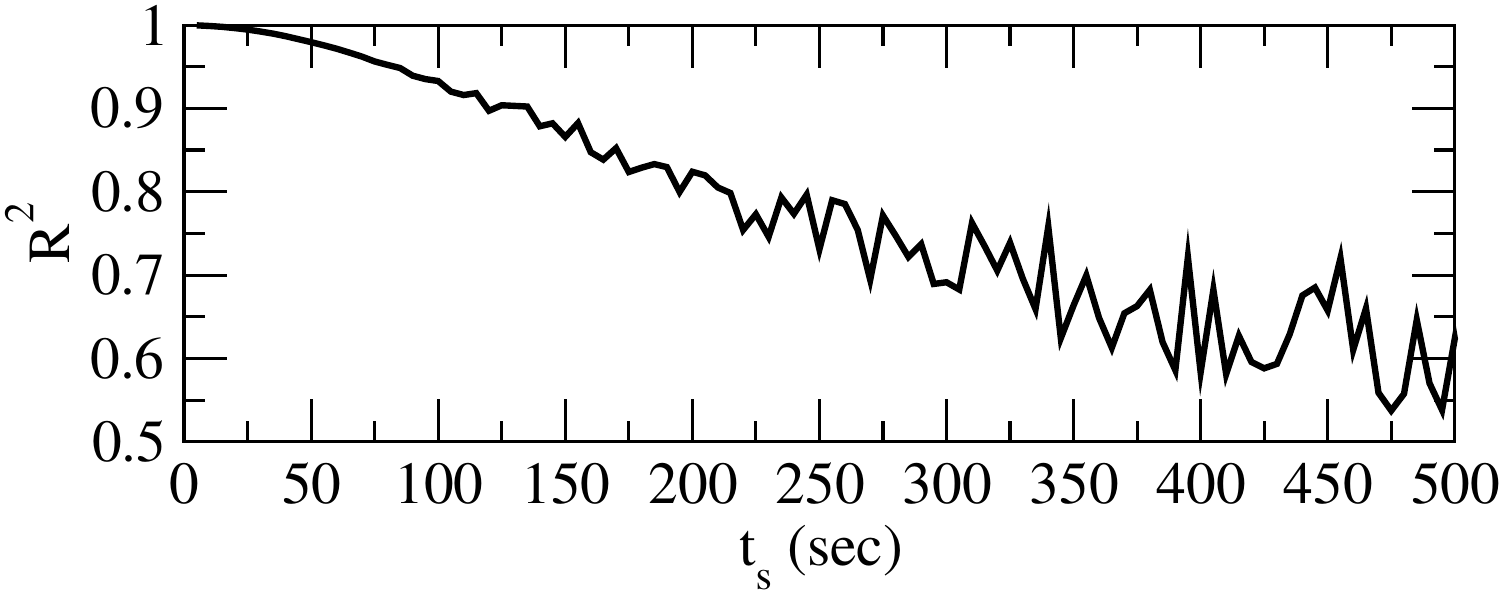}
 \caption{Average $R^2$ performance for increasing values of the
   sampling rate using 3 months of training and testing in the
 following month. Beyond the two minute sampling rate the $R^2$ drops
   below 0.9.  }\label{fig:sampling}
\end{figure}

Another important question is determining an acceptable sampling rate
on the electronic nose to be able to filter the humidity and the
temperature. We estimate the effect in terms of regression accuracy of
different sampling rates by computing the average $R^2$ values for all
the sensors modifying the sampling period from 5 to 500 seconds. In
Fig. \ref{fig:sampling}, we can see that beyond the 2 minute sampling
period, the filter performance drops below 0.9. Beyond this point, the
approximations made in the band-based model in equations
(\ref{eq:app}-\ref{prediction}) fail.

Faster sampling rates may still be required to implement for some
strategies that use sensor heater control in an active manner
\cite{herrero2015active} or in fast changing environments. However,
further work is still needed to consider highly ventilated scenarios
in which temperature and humidity change in time at the same rate as
the atmosphere chemical composition.. Comparatively, an empirical
approach can be found in \cite{piedrahita2014next}, where a similar
model is fitted to a linear dependence on temperature and humidity,
but not on the changes of the temperature and humidity.

\section{Impact on online discrimination of gas identity}
\label{sec:performance}

\begin{figure}[t!]
 \centering
 \includegraphics[width=1.0\textwidth]{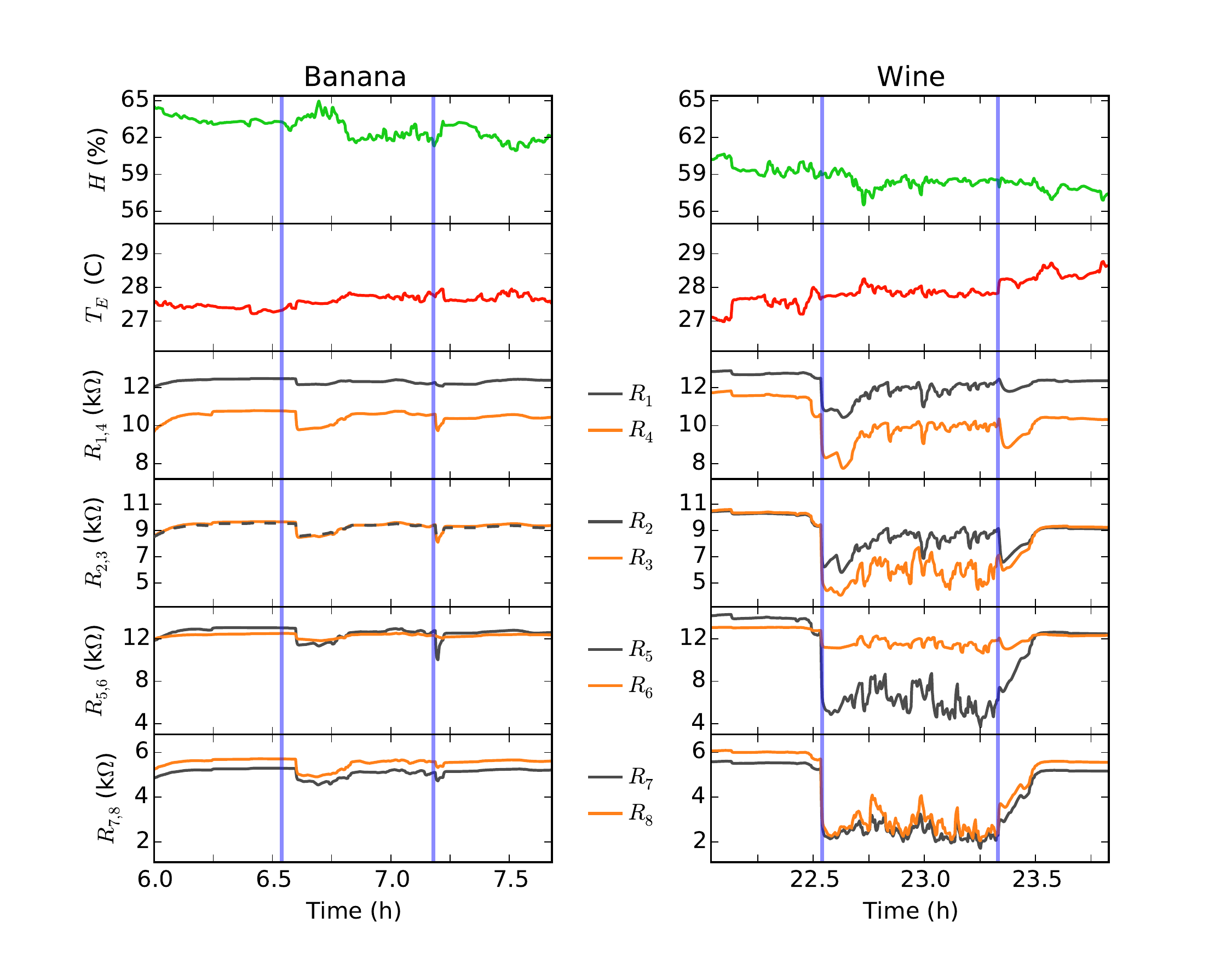}
 \caption{Example of response of all sensors due to the presentation
   of our stimuli: banana and wine. Sensors are indexed according to
   table \ref{tablesensorts}. Vertical blue lines delimit the period
   of time that the stimulus remained close to the electronic
   nose. These time series were recorded on September 22nd,
   2015. \label{fig:bananawineExample} }
\end{figure}

To investigate whether a predictive model can potentially benefit from
filtering temperature and humidity sensors, we constructed a data set
from recordings of two distinct stimuli: wine and banana
(Fig. \ref{fig:bananawineExample}). We compared the impact of using
the raw data and the filtered data in terms of classification
performance when discriminating among presence of banana, wine and
lack of stimulus (i.e., background activity). Signals recorded with
banana or wine evoked different responses in the sensors. In
particular, responses to banana were often weaker and returned to the
baseline activity much faster than those of wine (compare for instance
$R_4$ in Fig. \ref{fig:bananawineExample}).  Rather than using the
particular chemical signatures of compounds from bananas and wines,
our goal is to construct a model that learns to predict presence of
banana/wine based on the multivariate response of the sensors. The
chemical signature of bananas changes, for instance, as they ripen
\cite{llobet1999non}, and wine's signature depends on alcohol content
(ethanol), origin of the grape, among other factors
\cite{lozano2005classification,di1996electronic}. Thus, our approach
attempts at building a model that does not rely on wine type and
banana ripeness.

These data were collected over the course of 2 months by placing a
sample of either a banana or wine next to the electronic nose for a
period of time ranging from 10 minutes to 1 hour. Baseline signals
were taken from 2PM to 3PM to avoid additional noise due to home
activity. The time of the day when the stimulus was presented varied,
except between 12AM and 6AM. On total, our dataset comprises the time
series of 34 banana presentations, 36 wine presentations, and 30
baseline samples. To implement online discrimination, the data was
organized in moving windows with lengths of $10$ minutes. For
instance, for a presentation of length 60 min we create a total of 60
- 10 = 50 windows to be used during the classification.

To solve the classification problem, we used a nonlinear classifier
called Inhibitory Support Vector Machine (ISVM) \cite{isvm}, which, in
contrast to other multiclass SVM methods, is Bayes consistent for
three classes. ISVM is a particular case of the $\lambda$-SVM
classifier, a pointwise Fisher consistent multiclass classifier
\cite{rodriguez2015fisher}. ISVMs have been successfully applied to
arrays of electronic noses (identical to the one used in the present
paper) in controlled
conditions~\cite{rodriguez2014calibration,rodriguez2015fisher}, in
wind tunnel testing~\cite{vergara2013performance}, and for ethylene
discrimination in binary gas mixtures
\cite{fonollosa2014chemical}. Inspired by the learning mechanisms
present the insect brain \cite{201480:curropinion}, Inhibitory SVMs
use a large-margin classifier framework coupled to a mechanism of
mutual and unselective inhibition among classes. This mutual
inhibition creates a competition, from which only one class
emerges. The decision function of Inhibitory SVMs associated with the
$j$-th class and the input pattern $\vect{x}_i$ is defined as
$f_j(\vect{x_i}) = \langle \vect{w_j} , \Phi(\vect{x_i}) \rangle - \mu
\sum_{k=1}^L {\langle \vect{w_k} , \Phi(\vect{x_i}) \rangle}$, where
$L$ is the number of classes and $\mu$ scales how strong each class
will inhibit each other. If $\mu=0$, the decision function for
standard SVMs is recovered. It can be analytically shown that the
optimal value for $\mu$ is $1/L$. The predicted class of a data point
$\vect{x}_i$ is determined by the maximum among the decision functions
for each class: $y(\vect{x}_i)=\arg \max_j f_j(\vect{x}_i)$. Because
we used Radial Basis Functions (RBF) as the kernel of the inhibitory
SVM, our classifier had two meta-parameters: the soft margin
penalization $C$, and the inverse of the scale of the RBF function
$\gamma$. For more details about the ISVM model, see
\cite{isvm,rodriguez2015fisher}.

To evaluate the impact on discrimination performance due to decorrelating the
signals from temperature and humidity sensors, we tested $4$ different feature
sets: raw sensor time series (RS), raw sensor data with humidity and temperature
(RS,T,H), filtered data (FS) by decorrelating sensors using equation
\ref{eq:decorrmethod}, and raw sensor data with filtered sensor data (RS,FS). To
properly estimate the generalization ability of the model, we used standard
procedures in machine learning to evaluate the performance of our classifier
when discriminating samples not used for training the classifier
\cite{duda2012pattern}. We first divided our data set into two groups: a
training set with 4/5th of the experimental presentations, and a test set with
1/5th of the data. All moving windows associated with the same presentations
were kept in the same group. We used 4-fold cross-validation on the training set
to estimate the classifier meta-parameters ($C$ and $\gamma$). Using these
meta-parameters, we re-trained the model using the whole training set and, then,
assessed the performance using the test set. The range of values for the
meta-parameters explored during the 4-fold cross-validation in the training set
were $\gamma=\lbrace 0.5, 1, 5, 10, 50, 100\rbrace$, and $C=\lbrace 10^4, 10^5,
10^6, 10^7, 10^8, 10^9 \rbrace$. To obtain a good statistical estimate of the
classification accuracy, we re-shuffled our data and repeated this procedure
$50$ times, which was enough for the average and variance to converge.

Using the raw sensor data combined with the filtered signals (RS,FS)
improved significantly (Kolmogorov-Smirnov, $p<0.025$) the performance
in online discrimination (Table~\ref{tableresultsclass}). The raw
sensors data (RS) alone reached $76\%$ of accuracy, and including the
temperature and humidity information (RS,T,H) did not improve. This
shows that the additional features are likely redundant. Probably due
to loss of inter-dependencies among sensors (as anticipated in section
\ref{sec:performancemodel}), the filtered sensor data (FS) by itself
underperformed RS. Still, the model becomes more consistent, with
lower variance in its performance, than the models trained on (RS) and
(RS,T,H). Indeed, using both raw and filtered time series (RS,FS)
improved significantly the model performance and its
consistency. Thus, this experiment illustrates that temperature and
humidity filters can not only improve pattern recognition performance,
but they can also improve model stability, which is especially
challenging in chemical
sensing~\cite{romain2010long,padilla2010,carlo2011,Vergara2012,martinelli2013adaptive}.

\begin{table}[t]
  \centering
  \begin{tabular}{@{}|l|ccc|c|}
    \hline
    Feature set & Cross-validated accuracy &  Accuracy in test  & Std & p-value\\
    \hline
    RS & 78.5\% & 76.5\% & 6.8\% & $0.02^*$ \\
    RS,T,H & 73.3\% & 71.1\% & 6.8\% & $1\cdot10^{-12}$ $^{**}$ \\
    FS & 72.4\% &  71.2\% &   {\bf 4.8}\% & $2\cdot10^{-12}$ $^{**}$\\
    RS,FS & 82.6\% &  {\bf 80.9}\%  &  6.3\% & 1\\
    \hline \hline
  \end{tabular}
  \caption{Classification accuracies in four feature sets
    (abbreviations are defined in the text) derived from our dataset
    with three classes: wine, banana, and baseline activity.
    The meta-parameters of the final Inhibitory SVM model were selected
    as those with the best cross-validated accuracies in the training set
    (second column), and the generalization error of the final model was
    evaluated in the test set (third column). The standard deviation
    (\textit{std}) for the test dataset is estimated over $50$
    random partitions. Accuracy results from (RS,FS) are significantly
    different from all other feature sets (p-values from
    Kolmogorov-Smirnoff tests, $**$ passes at 1\%, $*$ passes at
    5\%).}
  \label{tableresultsclass}
\end{table}

\section{Conclusions}

Changes in humidity and temperature shape the responses of arrays of
MOX sensors, which in turn modifies nonlinearly chemical signatures of
different volatiles. Filtering changes in the sensor responses due to
changes in both humidity and temperature during sampling represents a
major improvement for complex machine learning and monitoring
tasks. We used a model based on semiconductor energy bands to express
the nonlinear dependence of sensor resistance with humidity and
temperature variations in an electronic nose. The model was designed
to fit in simpler micro-controllers, removing all possible
non-linearities up to second order in the change of humidity and
temperature, envisioning applications to cost-efficient devices. We
found that the most dominant terms are the change in humidity, the
quadratic term of the change in humidity, and the correlated
variations of humidity and temperature. We showed that the model
provides robust corrections to the distortions caused by environmental
changes. Therefore, our level of approximation on the semiconductor
energy band is an inexpensive solution for applications in online and
continuous home monitoring using chemical sensors.

Specifically, the coefficient of determination $R^2$ of our model when
fitted to all the 537 days of sampling is remarkably close to
$100$\%. The model predicts a particular dependence between two of the
coefficients that is consistently verified in all the tested
sensors. We also showed that the maximum sampling period to obtain a
reliable filter of humidity and temperature is of the order of 1
minute. The accuracy achieved with faster sampling rates provides
small gains, and it would require some overhead in wireless
communication when the corrections are done at the base
station. Additionally, 3-month training window was selected to ensure
that $R^2$ is larger than $90$\% for all sensors and throughout the
whole year. With 3 months, the training dataset likely included enough
number of training examples (events and background) while the effect
of long-term drift in the sensors was still weak to degrade the
trained models. Previous work using similar sensing units showed that
models trained in two-month windows keep high accuracy during the
following two months \cite{vergara2012chemical}. Stability could
probably be improved further if one selects longer training windows or
by coupling our strategy with already proposed strategies to
counteract long-term sensor drift
\cite{vergara2012chemical,padilla2010,ziyatdinov2010drift}.

We verified empirically the benefits of decorrelating humidity and
temperature from the sensors' response by applying it to a task of gas
discrimination. We recorded the response of the sensors when presented
with either a banana or glass of wine. Then, we used a
Bayes-consistent classifier \cite{rodriguez2015fisher,isvm} to
discriminate between the presence of banana, presence of wine, and
baseline activity. To compare the performance of the classifier with
and without the decorrelation of humidity-temperature, four different
subsets of data were created by combining raw sensor responses,
filtered sensor data, and temperature and humidity. Experimental
results show that including the filtered data in the classification
model improves not only the discrimination capability of the model,
but also its stability.

In summary, we have shown that simultaneous recordings of the humidity
and the temperature can be used to help extracting relevant chemical
signatures. The online decorrelation model proposed in this work was
designed for online operation even in the simpler micro-controllers
available in the market, which is essential for cost-efficient
devices. Additionally, humidity sensors are extremely appealing due to
a high correlation between humidity levels and human perception of air
quality \cite{wolkoff2007dichotomy,fang1998impact}. Thus, when
combined with other techniques
\cite{fonollosa2014human,rodriguez2014calibration,reservoir,fonollosa2014chemical,diamond2016classifying,mosqueiro2016CISS},
our model is likely to significantly enhance the performance of
chemical detection systems, as for instance of home monitoring
tasks. Our contribution thus emphasizes the importance of simultaneous
recordings of humidity and temperature, and that their use is
computationally amenable in sensor boards using low-energy
micro-controllers.

\section*{Acknowledgments}

This work has been supported by the California Institute for
Telecommunications and Information Technology (CALIT2) under Grant
Number 2014CSRO 136.  JF acknowledges the support of the Marie Curie
Actions and the Agency for Business Competitiveness of the Government
of Catalonia (ACCI\'O) for the grant TECSPR15-1-0031; and the Spanish
MINECO program, grant TEC2014-59229-R (SIGVOL). RH, TM, and IR-L
acknowledge the partial support by 3\textordfeminine~Convocatoria de
Proyectos de Cooperacion Interuniversitaria UAM--Banco Santander con
EEUU. NR would like to acknowledge partial support by ONR grant
N000141612252. TM acknowledges CNPq grant 234817/2014- 3 for partial
support. We are also thankful to Flavia Huerta who collected data
examples during the summer of 2015.

\bibliographystyle{unsrt}
\bibliography{master}

\end{document}